\documentclass{article}
\usepackage{arxiv}

\usepackage[utf8]{inputenc} 
\usepackage[T1]{fontenc}    
\usepackage{hyperref}       
\usepackage{url}            
\usepackage{booktabs}       
\usepackage{amsfonts}       
\usepackage{nicefrac}       
\usepackage{microtype}      
\usepackage{lipsum}		
\usepackage{graphicx}
\usepackage{cite}

\usepackage{array}
\usepackage{amssymb, amsmath, amsthm}
\usepackage{graphicx}
\usepackage{lmodern,url}
\usepackage{makecell} 
\usepackage{cancel}
\usepackage{multirow}
\usepackage{microtype}
\usepackage{lineno}
\usepackage{xspace}
\usepackage{xcolor}
\usepackage{siunitx}
\usepackage{todonotes}
\usepackage{booktabs}
\usepackage[ruled,vlined]{algorithm2e}
\usepackage{optidef}
\usepackage{mathdots}
\usepackage{subcaption}
\usepackage{csquotes}
\usepackage{caption}
\newcommand{\dis}{\displaystyle}

\newcommand{\figref}[1]{Fig.~\ref{#1}}
\newcommand{\tabref}[1]{\tablename~\ref{#1}}
\graphicspath{{Figures/}}

\usepackage{amsmath}
\usepackage{tikz}
\usetikzlibrary{fit}
\usetikzlibrary{positioning}
\tikzset{%
  highlight/.style={rectangle,rounded corners,fill=red!15,draw,fill opacity=0.25,thick,inner sep=0pt}
}

\definecolor{mygreen}{RGB}{160, 242,182}

\newcommand{\xunderbrace}[2][\vphantom{\dfrac{A}{A}}]{\underbrace{#1#2}}
\definecolor{mypurple}{RGB}{236, 223, 234}

\newcommand{\RtH}{\ensuremath{R_t}\xspace}
\newcommand{\RtBase}{\ensuremath{R_t^{\rm NPI}}\xspace}
\newcommand{\Ieff}{\ensuremath{I_{\rm eff}}\xspace}
\newcommand{\EB}{\ensuremath{E^B}\xspace}
\newcommand{\IB}{\ensuremath{I^B}\xspace}
\newcommand{\virload}{\ensuremath{\sigma}\xspace}
\newcommand{\ICU}{\ensuremath{{\rm ICU}}\xspace}
\newcommand{\avICU}{\ensuremath{H}\xspace}
\newcommand{\latRate}{\ensuremath{\rho}\xspace}
\newcommand{\fracImmun}{\ensuremath{\eta}\xspace}

\newcommand{\avProtection}{\ensuremath{\kappa}\xspace}

\allowdisplaybreaks

\title{The winter dilemma}

\usepackage{authblk}

\author[1$\dagger$]{Sebastian Contreras}
\author[1$\dagger$]{Philipp Dönges}
\author[1$\dagger$]{Joel Wagner}
\author[1$\dagger$]{Simon Bauer}
\author[1]{Sebastian B. Mohr}
\author[1]{Emil N. Iftekhar}
\author[2]{Mirjam Kretzschmar}
\author[3]{Michael M\"as}
\author[4]{Kai Nagel}
\author[5]{Andr\'e Calero Valdez}
\author[1,6*]{Viola Priesemann}

\affil[1]{Max Planck Institute for Dynamics and Self-Organization, G\"ottingen, Germany.}
\affil[2]{University Medical Center Utrecht, Utrecht, The Netherlands.}
\affil[3]{Karlsruhe Institute of Technology, Karlsruhe, Germany.}
\affil[4]{Technische Universität Berlin, Berlin, Germany.}
\affil[5]{RWTH Aachen University, Aachen, Germany.}
\affil[6]{Institute for the Dynamics of Complex Systems, University of G\"ottingen,  G\"ottingen, Germany.}
\affil[ ]{{$*$} Corresponding Author: Viola Priesemann (viola.priesemann@ds.mpg.de)}
\affil[ ]{{$\dagger$} These authors contributed equally}

\date{}

\renewcommand{\headeright}{ }
\renewcommand{\undertitle}{ }

\begin{document}
\maketitle

\begin{abstract} 
With winter coming in the northern hemisphere, disadvantageous seasonality of SARS-CoV-2 requires high immunity levels in the population or increasing non-pharmaceutical interventions (NPIs), compared to summer. Otherwise intensive care units (ICUs) might fill up.
However, compliance with mandatory NPIs, vaccine uptake, and individual protective measures depend on individuals' opinions and behavior. Opinions, in turn, depend on information, e.g.,  about vaccine safety or current infection levels. Therefore, understanding how information about the pandemic affects its spread through the modulation of voluntary protection-seeking behaviors is crucial for better preparedness this winter and for future crises.
\end{abstract}

\begin{figure}[!ht]
    \centering
    \includegraphics[width=6.5in]{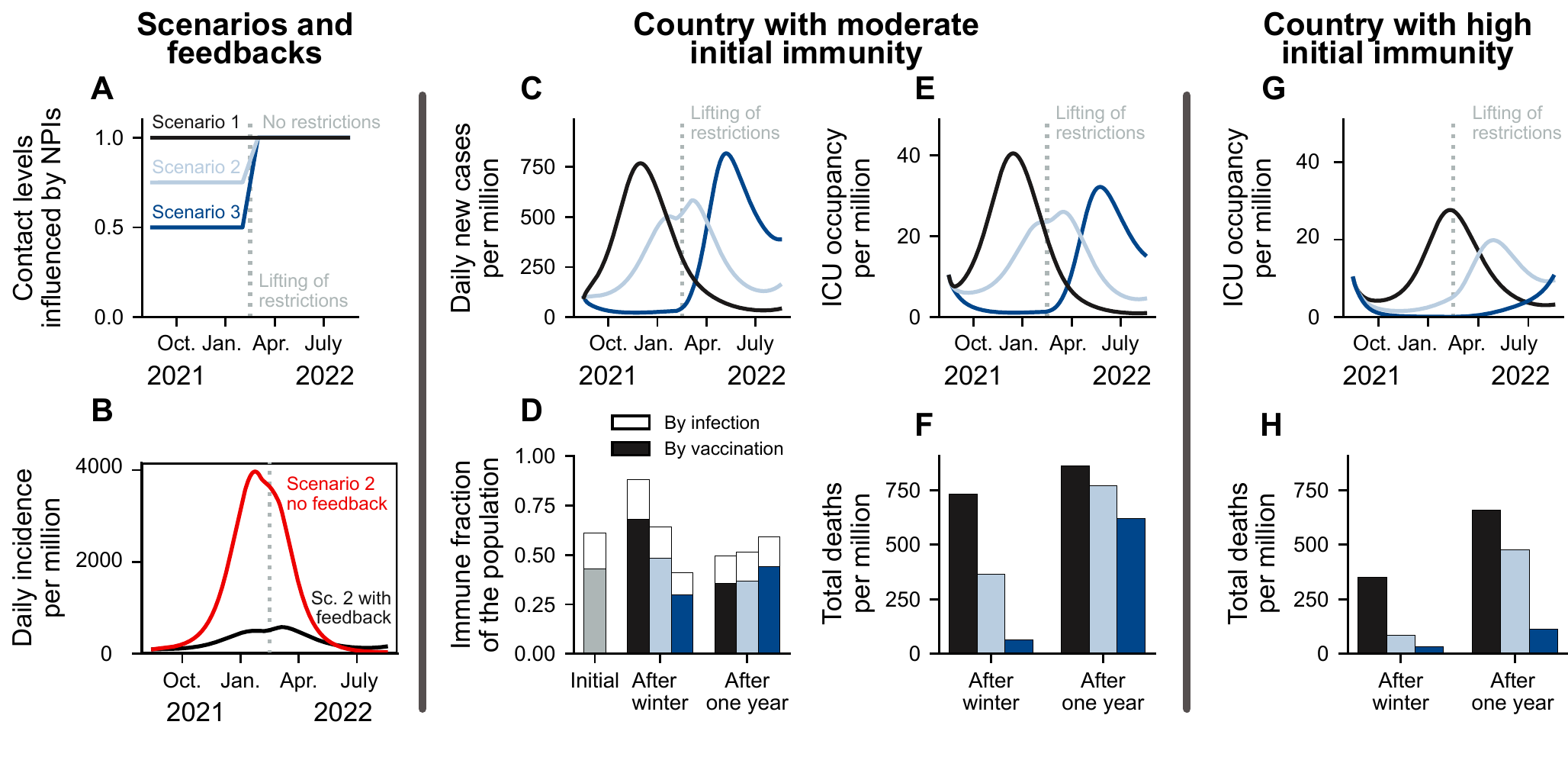}
    \caption{%
        \textbf{COVID-19 restrictions planning through winter: a long-term dilemma.} The interplay of non-pharmaceutical interventions (NPI), that are assumed to be sustained through winter 2021/2022, together with people's protection-seeking behavior will determine case numbers and ICU occupancy over winter and beyond. \textbf{A, B:} We explored three scenarios of mandatory NPI stringency in winter (\textbf{A}), and including the feedback loop between information and disease spread (\textbf{B}). We use as example countries with moderate (\textbf{C--F}) and high (\textbf{G--H}) levels of vaccine-induced immunity. \textbf{C--F:} Scenario 1: having no restrictions causes a steep increase in case numbers and ICU occupancy that triggers protection-seeking behavior among the population. In this situation, the self-regulation of contacts, growing vaccine uptake, and higher rates of natural immunization would contribute to stabilizing case numbers (\textbf{C, D}), bearing, however, high mortality and morbidity in winter (\textbf{F}). Scenario 2: Maintaining mild restrictions would curb the overwhelming of ICUs while motivating higher vaccination  and natural immunity rates. Scenario 3: Maintaining moderate restrictions throughout winter will minimize COVID-19 cases and hospitalizations in winter, generating a shared perception of safety across the population. However, low vaccine uptake and rates of naturally acquired immunity through winter together with waned immunity will cause a severe rebound wave when restrictions are completely lifted in March (\textbf{D--F}). \textbf{G, H:} A country facing winter with higher vaccine coverage will not face the dilemma, but might require additional measures to prevent a larger seasonal wave in the subsequent winter. Note that in (\textbf{D}) the upper white fraction (immune by infection) also includes those who had been infected and then also vaccinated.
        }
    \label{fig:Figure_1}
\end{figure}

Protection-seeking behavior increases when individuals perceive high personal risks of infection, which increases when COVID-19 incidence and ICU occupancy rise. This interdependency between information and behavior generates a dilemma for the coming winter. On the one hand, maintaining moderate levels of NPIs to keep the reproduction number $R$ low, implies decreasing COVID-19 incidence---in turn diminishing incentives to reduce contacts or get vaccinated; thereby, one risks a severe wave as soon as restrictions are lifted (especially considering waning immunity from those immunized long ago). On the other hand, relaxing restrictions more than current immunity levels allow can lead to excess morbidity and mortality (Fig. \ref{fig:Figure_1}) ~\cite{bauer2021relaxing, contreras2021risking}. 

To demonstrate the wickedness of the winter dilemma, we use a standard susceptible-exposed-infected-recovered (SEIR) model with explicit compartments for fatalities, ICUs, and vaccination (first time and booster vaccines), and also waning immunity and seasonality \cite{contreras2021winter,teslya2020impact}. To account for behavioral change induced by perceived risk of infection, we include a feedback loop between information on ICU occupancy and the level of contacts, i.e., the reproduction number, and vaccination willingness \cite{contreras2021winter}. Explicitly, we assume that increases in ICU occupancy i)~decrease the spreading rate of COVID-19, accounting for protection-seeking behavior and voluntary reduction of mobility \cite{nouvellet2021reduction,druckman2021affective}, and ii)~increase vaccine acceptance among hesitant individuals \cite{salali2021effective}. 

Importantly, if not including these behavioral feedback loops, then incidence and hospitalization might get extremely high (Fig. \ref{fig:Figure_1} B). 
With including the feedback-loop, we analyzed three scenarios of mandatory NPIs during winter: 1)~immediately lifting all NPIs, 2)~maintaining mild NPIs, and 3)~maintaining moderate NPIs to sustain low case numbers (cf.~Fig\ref{fig:Figure_1}A). 

Without any NPIs, and at moderate immunity at the start of winter, we expect a steep rise in case numbers and hospitalization (Fig~\ref{fig:Figure_1}C--F, black lines). As a consequence, individuals are expected to voluntarily reduce their contacts and are more inclined to accept a vaccine offer (Fig~\ref{fig:Figure_1}D). However, this surge will increase morbidity and mortality because the effect of vaccination is not instantaneous. The opposite corner scenario---sustaining moderate NPIs and low case numbers---might lead to low COVID-19 incidence during winter but risks a rebound wave in spring (Fig~\ref{fig:Figure_1}, blue). This is because the low incidence during winter may imply i)~low natural immunity, ii)~lacking incentives for vaccination, and iii)~lower chances of refreshing immune memory upon re-exposure to the virus \cite{brown2021original}. The resulting low immunity levels (cf. Fig~\ref{fig:Figure_1} D) can then fuel a high rebound wave in spring. Similar rebound waves have been observed for other seasonal respiratory viruses \cite{gomez2021uncertain}. Countries starting winter with higher immunity levels among the population (cf. Fig.~\ref{fig:Figure_1}G, H) will not observe a steep increase on case numbers. However, as before, waning immunity can postpone the wave to the next winter.

In sum, the way governments approach winter will shape long-term COVID-19 transmission dynamics and thus determine i)~the probability of having an off-seasonal COVID-19 wave when lifting NPIs, ii)~the magnitude of the self-regulation effect induced by the information-behavior feedback loop, and iii)~how we will reach appropriate immunity levels to transit from epidemicity to endemicity smoothly. 

The simple solution to the winter dilemma is obvious: A higher vaccination rate and boostering (>80 or 90 \%, depending on age), especially among the elderly could avoid either wave even if NPIs are abolished \cite{contreras2021winter}. However, at low to intermediate vaccination, the solution is not obvious; in any scenario, we expect moderate to high burden to the health system (Fig~\ref{fig:Figure_1}F).
Thus most importantly, the challenge for authorities is to find ways to engage individuals with vaccination programs without requiring high case numbers for that. Here, clear communication and trust continues to be essential  \cite{iftekhar2021}.


\section*{Data availability}
Model and equations are presented in \cite{contreras2021winter}. Source code for data generation and analysis is available online on GitHub \url{https://github.com/Priesemann-Group/covid19_winter_dilemma}.

\section*{Acknowledgments} 
Authors with affiliation received support from the Max-Planck-Society. SB and SBM were financially supported by the German Federal Ministry of Education and Research (BMBF) as part of the Network University Medicine (NUM), project egePan, funding code: 01KX2021.


\newpage

\renewcommand{\thefigure}{S\arabic{figure}}
\renewcommand{\figurename}{Supplementary~Figure}
\setcounter{figure}{0}
\renewcommand{\thetable}{S\arabic{table}}

\setcounter{table}{0}
\renewcommand{\theequation}{\arabic{equation}}
\setcounter{equation}{0}
\renewcommand{\thesection}{S\arabic{section}}
\setcounter{section}{0}
\section*{Supplementary Material}

\section{Model}
\begin{figure}[!h]
    \centering
    \includegraphics[width=5.5in]{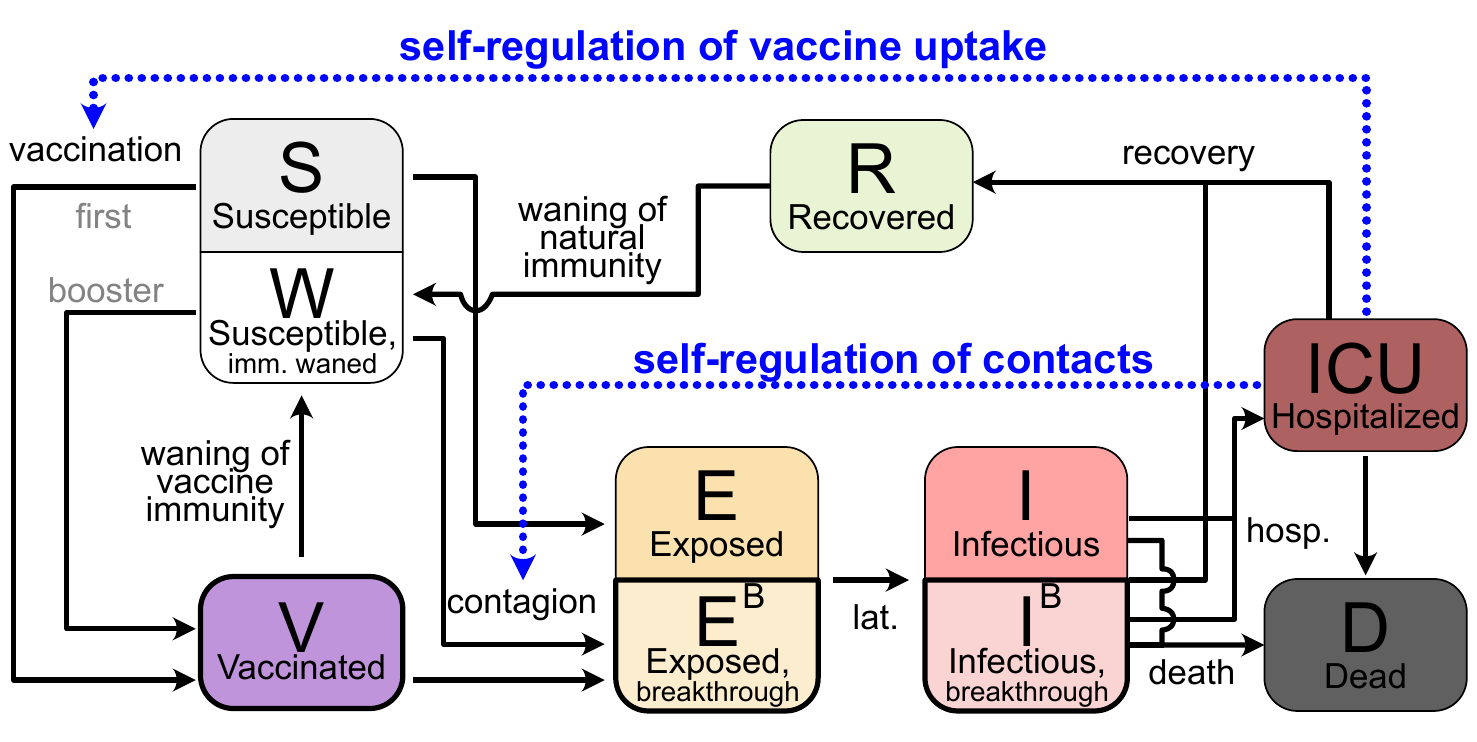}
    \caption{%
        \textbf{Figure S1: Compartmental model with behavioral feedback loops between hospital occupancy and spreading rate and vaccination willingness.} Transition rates are listed in Table S1, but omitted in the figure for clarity purposes.
        }
    \label{fig:Figure_S1}
\end{figure}

We model the spreading dynamics of SARS-CoV-2 following a mean-field SEIRD-ICU deterministic formalism through a system of differential equations (\figref{fig:Figure_S1}). Our model incorporates disease-spreading dynamics, ICU stays, and the roll-out of a single-dose vaccine (representing also the  two doses of most COVID-19 vaccines). Both vaccine-induced and naturally-acquired immunity wane over time, but vaccine-induced immunity wanes faster. In our model, susceptible individuals can acquire the virus from infected individuals and subsequently progress to the exposed ($S\rightarrow E$) and infectious ($E\rightarrow I$) compartments. We assume that vaccines offer no perfect sterile immunity and that a fraction of vaccinated people are infected upon contact with the infectious groups, i.e., we implement breakthrough infections. 
In contrast to susceptible individuals $S$, if infected, the individuals  with waned immunity $W$ and those vaccinated  move to specific exposed and infectious compartments ($V, W\rightarrow \EB \rightarrow \IB$). 
This is to implement that  the breakthrough infected still have a moderate protection against a severe course of the disease, i.e., have reduced probabilities to go to ICU or die.

Individuals exposed to the virus ($E, \EB$) progress from the exposed to the infectious compartments ($I, \IB$) at a rate \latRate. The infectious compartments have three different possible transitions: i) direct recovery ($I, \IB \rightarrow R$) with rate $\gamma$, ii) progression to ICU ($I, \IB \rightarrow \ICU$) with rate $\delta$ (reduced by $(1-\kappa)$ for $\IB$) or iii) direct death ($I, \IB \rightarrow D$) with rate $\theta$ (reduced by $(1-\kappa)$ for $\IB$). Individuals receiving ICU treatment recover either at a rate $\gamma_{\ICU}$ ($\ICU\rightarrow R$) or die at a rate $\theta_\ICU$ ($\ICU\rightarrow D$). All parameters are listed in Tab: \ref{tab:Parametros}.

Another important property of this model is the self-regulation of contacts and vaccine acceptance that influences the disease and vaccination dynamics based on the current and past ICU occupancy (blue arrows in \figref{fig:Figure_S1}). 

The severity of COVID-19 infections strongly depends on age. Therefore, we used age-dependent rates $k_i$ for the transition rates to $\ICU$ and $D$ compartments, as in \cite{bauer2021relaxing}. We calculate the overall transition rate as $\bar k = \sum_i w_i k_i$, where $i$ denotes the age group and $w_i = \frac{M_i}{M}$ is the fraction of age group $i$ of the total population. 

\subsection{Reproduction Number}

The reproduction number $R_t$ includes (i) the effects of mandatory non-pharmaceutical interventions, (ii) individuals self-regulating their contacts based on perceived risk, and (iii) seasonality. Each is represented by a multiplicative factor on the basic reproduction number $R_0$, i.e., the total number of offspring infections that a single case would generate in a fully susceptible population without any restrictions.

First, what we call the "NPI-related reproduction number" \RtBase reflects the $R_0$ with a potential reduction due to mandatory NPIs over the winter. For \RtBase, we chose three scenarios (cf. Fig.~\ref{fig:Figure_1}): 
The immediate lifting of all restrictions (high \RtBase), weak restrictions over winter (moderate \RtBase) and moderate restrictions over winter (low \RtBase). Note that we do not deem strong restrictions necessary over winter. For all scenarios, we assumed that in March 2022 all restrictions will be lifted. Easy-to-follow measures such as improved hygiene might still be kept in place which results in a small reduction of the basic reproduction number $R_0$. The implementation and abolishment of NPIs is modeled by a linear decrease or increase, respectively, in \RtBase that lasts four weeks. 

Second, to implement that each individual has the freedom to adapt his or her behavior in accordance to perceived risk, we implemented a further parameter reducing the $R_t$. This factor depends on the past ICU occupancy $\avICU_R$. In detail, to implement this behavioral feedback-loop, we use an exponentially decaying term (see sec. \ref{sec:memoryICU}), where the decay depends on $\avICU_R$ and a sensitivity constant $\alpha_R$. 

Third, seasonality is modeled by a time-dependent sinusoidal modulation factor $\Gamma(t)$ that depends on the sensitivity $\mu$ and the day with the highest effect on seasonality $d_\mu$, which for our purpose can be set to zero, corresponding to January 1 \cite{Gavenciak2021seasonality}. The full time- and behavior-dependent reproduction number is then given by 

\begin{align}
    R_t(\avICU_R, t) & = \RtBase \exp\left(-\alpha_R \avICU_R\right) \frac{\Gamma(t)}{\Gamma(360-d_0)}, \label{eq:Rt}\\ 
    \Gamma(t)      &= 1+\mu\cos\left(2\pi\frac{t+d_0-d_\mu}{360}\right) \label{eq:seasonality}
\end{align}
where $\frac{1}{\Gamma(360-d_0)}$ is for normalization such that seasonality only decreases $R_t$, i.e., neglecting the behavior term, $\RtBase$ corresponds to the peak value in winter. For simplicity in our model, one month has 30 days and a full year thus 360 days which does not affect the results on our time horizon. 

\subsubsection{Memory on perceived danger}
\label{sec:memoryICU}
Perceived danger for the individual, transmitted by e.g., mass media or affected acquaintances, depends on ICU not only at the present moment but also on the past. That way, self-regulation of contacts and vaccine uptake is a function of the past development of the ICU occupancy called $\avICU$. We assume that the memory of past ICU development is smooth, meaning that past ICU occupancies are remembered less and less as time passes. To incorporate this into our model we calculate the convolution of the ICU with a gamma distribution, effectively "weighting" the past development of ICU. That way, ICU occupancy a few days ago is "remembered" more and thus influences people's behavior at the present moment more than the ICU occupancy that lies further in the past. That way, the reproduction number becomes dependent on $H_R(t)$ via

\begin{equation}
    H_R(t) := \ICU \ast \mathcal{G}_{p_R,b_R} = \int_{-\infty}^{t} \text{dt}'\, \ICU(t')\mathcal{G}_{p_R, b_R}(-t'+t)\,,
\end{equation}
where the arguments of the gamma distribution are set to $p_R=0.7$ and $b_R=4$. 

Time memory for vaccination willingness is assumed to work in the same way but with different Gamma distributions. First of all, there is a delay between the decision to be vaccinated and the onset of immunity. Secondly, vaccination willingness is assumed to depend on a longer time interval of the ICU occupancy. Combined, it translates into a Gamma distribution that is shifted in time and looks flatter which is characterized by the parameters $\tau_u$, $\tau_w$ and $b_v=14$:

\begin{equation}
    H_{u,w}(t) := \ICU \ast \mathcal{G}_{p=1, b_{v}} = \int_{-\infty}^{t} \text{dt}'\, \ICU(t')\mathcal{G}_{p=1, b_{v}}(-t'-\tau_{u,w}+t).
\end{equation}

The subscripts $_u$ and $_w$ indicate first and booster doses respectively.
The parameter $\tau_u$ is larger than $\tau_w$ because we include the delay of around 6 weeks for most vaccines that need two doses. Booster doses are usually only a single dose so $\tau_w$ is just the delay between administration of the dose and onset of immunity which we assume to be 2 weeks. For the initial conditions of $\avICU$ and $\avICU_R$ we set $\ICU$ to a constant $\ICU(t<0)=\ICU(t=0)$ in the past. This simplification affects the results only negligibly for a short initial time.

\subsubsection{Waning Immunity}

Our model includes two types of immunity: immunity as a result of vaccination and immunity as a result of natural infection. In both cases, immunity wanes over time although it is believed that natural immunity lasts longer and thus has a lower waning rate. On average, vaccine-induced immunity wanes after $(\Omega_v^{\rm base})^{-1}$ months and naturally-acquired immunity after $(\Omega_n^{\rm base})^{-1}$ months. Furthermore, we assume that immunized individuals can "refresh" their immune memory upon contact with the virus which translates into infection level- and $R_t$-dependent waning rates $\Omega_{v,n}(\Ieff, R_t)$, where the effective incidence $\Ieff$ corresponds to the total size of the infectious pools $I$ and $\IB$ but acknowledges reduced virulence of breakthrough infections (see sec. \ref{sec:vaccinationeffects}). Furthermore an influx of $\Psi$ was added to account for infections from abroad. In the limit of high infection levels, the waning rate should converge to zero and in the limit ($\Ieff \rightarrow 0$), where no refreshing happens, it should be at its base value $\Omega_{v,n}^{\rm base}$. Using a logistic function that meets these requirements and decreases linearly for low infection levels, we can express the waning immunity as a function of $\Ieff$, the reproduction number $R_t$, and form parameters $c_v$ and $c_n$:

\begin{align}
    \Omega_v(\Ieff, R_t) & = 2\Omega_v^{\rm base}\left( 1- \frac{ 1}{1+ \text{exp}\left( -c_v R_t \Ieff \right)} \right)\,,\\
    \Omega_n(\Ieff, R_t) & = 2\Omega_n^{\rm base}\left( 1- \frac{ 1}{1+ \text{exp}\left( -c_n R_t \Ieff \right)} \right)\,.
\end{align}

If it holds that $\Ieff R_t =\frac{1}{c_{v,n}}$, we get the approximation $\Omega_{v,n} \approx \Omega^{\rm base}/2$ so we can get an estimate for $c_{v,n}$: An incidence of $\Ieff =\frac{1}{R_t c_{v,n}}$ corresponds to the case when the rate of waning immunity is halved, meaning that every second individual had his immunity refreshed in a given time frame. To find $c_{v,n}$ we consider the incidence necessary such that, in this given time frame, half of the population was infected. Using that a typical infection lasts $\mathcal{O}(\frac{1}{\gamma + \delta +\theta})$ days, the incidence at which after a certain amount of time $T$ half the population was infected is $I = \frac{M}{2}\frac{1}{\gamma+\delta+\theta}\frac{1}{T}$. Because every individual on average refreshes the immunity of $R_t$ individuals we divide by $R_t$ and set this equal to $\Ieff =\frac{1}{R_t c}$. The time frame should be the waning immunity time frame $T=\left(\Omega_{v,n}^{\rm base}\right)^{-1}$. Thus, we can obtain an estimate for $c_{v,n}$ as

\begin{equation}
    c_{v,n} \approx \frac{2}{M}\frac{\left(\gamma + \delta +\theta \right) }{\Omega_{v,n}^{\rm base}}\,.
    \label{eq: waningparameter}
\end{equation}

\subsection{Model Equations}

The combined contributions of the infection-spreading and vaccination dynamics are represented by the set of equations below. The time evolution of our model is then completely determined by the initial conditions of the system. The first-order transition rates between compartments are given by the probability for an individual to undergo this transition divided by the average transition time e.g., the recovery rate $\gamma$ is the probability that an individual recovers from the disease divided by how the time span of the recovery process. Note that in principle $\gamma$ should be different for the $I$ and $\IB$ compartment, as the probability to recover is larger for breakthrough infections. We neglect this difference as it is negligible within the margin of error since the probability to recover is close to 1 in both cases.

\begin{align}
&\frac{d S}{dt} & = & -\xunderbrace{\gamma R_t(\avICU_R)\frac{S}{M}\Ieff}_{\text{unvaccinated infections}} &-& \xunderbrace{\phi(\avICU_u)M}_{\text{first vaccinations}} & &  \label{eq:dSdt} \\
& \frac{dV}{dt} & = & -\xunderbrace{\left(1-\eta\right)\gamma R_t(\avICU_R)\frac{V}{M} \Ieff}_{\text{breakthrough infections}} &+& \xunderbrace{\left(\phi(\avICU_u)+\varphi(\avICU_w)\right)M}_{\text{vaccinations }} &-&  \xunderbrace{\Omega_v V}_{\text{waning vaccine immunity}} \label{eq:dVdt} \\
& \frac{dW}{dt} & = & -\xunderbrace{\gamma R_t(\avICU_R)\frac{W}{M} \Ieff}_{\text{waned infections}} &-& \xunderbrace{\varphi(\avICU_w)M}_{\text{booster vaccinations}}  &+& \xunderbrace{\Omega_v V
+\Omega_n R}_{\text{waning immunity}} \label{eq:dWdt}\\
& \frac{dE}{dt} & = & \xunderbrace{\gamma R_t(\avICU_R)\frac{S}{M} \Ieff}_{\text{unvaccinated exposed}}   &-&\xunderbrace{\latRate E}_{\text{end of latency}} & &  \label{eq:dEdt}\\
&\frac{d\EB}{dt} & = & \xunderbrace{\gamma R_t(\avICU_R)\frac{\left(1-\eta\right)V+W}{M} \Ieff}_{\text{vaccinated and waned exposed}}  &-&\xunderbrace{ \latRate \EB}_{\text{end of latency}}  & &  \label{eq:dEBdt}\\ 
&\frac{dI}{dt} & = & \xunderbrace{\latRate E}_{\text{start of infectiousness}}   &-&\xunderbrace{ \left(\gamma + \delta + \theta\right) I}_{\rightarrow\text{ICU, death and recovery}} & &   \label{eq:dIdt}\\ 
&\frac{d\IB}{dt} & = & \xunderbrace{\latRate \EB}_{\text{start of infectiousness}}   &-&\xunderbrace{ \left(\gamma + (\delta+\theta)(1-\avProtection)\right) \IB}_{\rightarrow\text{ICU, death (reduced) and recovery}}  & &\label{eq:dIBdt}\\ 
&\frac{d\ICU}{dt} & = & \xunderbrace{\delta \left(I + (1-\avProtection)\IB\right)}_{\text{transition to ICU}}  &-&\xunderbrace{ (\gamma_\ICU+\theta_\ICU) \ICU}_{\text{recovery from ICU}}  & & \label{eq:dICUdt}\\
&\frac{dD}{dt} & = & \xunderbrace{\theta \left(I + (1-\avProtection)\IB\right)}_{\text{death without ICU}}   &+&\xunderbrace{ \theta_\ICU \ICU}_{\text{death in ICU}}  & & \label{eq:dDdt}\\
&\frac{dR}{dt} & = & \xunderbrace{\gamma \left(I+\IB\right)}_{\text{direct recovery}}  &+  & \xunderbrace{\gamma_\ICU \ICU}_{\text{recovery from ICU}} &-&\xunderbrace{ \Omega_n R}_{\text{waning natural immunity}}  \label{eq:dRdt}\\
&\frac{du^{\rm current}}{dt} & = & \xunderbrace{M\phi(\avICU_u)}_{\text{current first vaccinations}}  & &  && \label{eq:dIUCdt}\\ 
&\frac{dw^{\rm current}}{dt} & = & \xunderbrace{M\varphi(\avICU_w)}_{\text{current booster vaccinations}}  & &  && \label{eq:dIWCdt}\\ 
&\Ieff & = & \xunderbrace{\left(I+\sigma\IB+ \Psi\right)}_{\text{effective incidence}}  & &  && \label{eq:Ieff} 
\end{align}

\begin{table}[htp]\caption{\textbf{Model parameters.} The range column either describes the range of values used in the various scenarios.}
\label{tab:Parametros}
\centering
\begin{tabular}{l p{5.5cm} lll p{2cm}}\toprule
Parameter           & Meaning                          & \makecell[l]{Value \\ (default)}   & \makecell[l]{Range\\ }         & Units             &   Source  \\\midrule
$R_0$   & Basic reproduction number & $5.0$                     & --  & \SI{}{-}        & \cite{bauer2021relaxing,davies_estimated_2021,orostica2021mutational} \\
$\RtBase$   & NPI-related rep. number (gross) & $\{2.0,\,3.5,\,5.0\}$                     & 0--5  & \SI{}{-}        & \cite{bauer2021relaxing,davies_estimated_2021,orostica2021mutational} \\
$\fracImmun$        & Vaccine eff. against transmission    & 0.75                & 0.6--0.85  & \SI{}{-}        & \cite{levine2021decreased,mallapaty_can_2021,hall_effectiveness_2021,abu2021effectiveness}\\
$\kappa_\mathrm{obs}$       & Observed vaccine eff. against severe disease  & 0.95 & 0.75--0.98  & \SI{}{-}        & \cite{mahase_covid-19_2021, dagan_bnt162b2_2021,bar2021protection, abu2021effectiveness}\\
$\kappa$       & Vaccine eff. against severe disease  & 0.8  & \SI{}{-}        & Eq.~\ref{eq:kappa}\\
$\virload$          & Relative virulence of vaccinated to unvaccinated individuals    & 0.5        & 0.5 -- 1  & \SI{}{-}  &  \cite{harris2021impact,martinez2021effectiveness}\\
$\tau_u, \tau_w$              & Memory time of the ICU capacity and delay to immunization    & 2, 6     &  -- & \SI{}{weeks}  & Assumed \\
$\rho$            & Latency rate                        & 0.25      & -- & \SI{}{day^{-1}}  & \cite{bar2020science, li2020substantial}\\
$\gamma$            & Recovery rate                         & 0.1      & 0.088 -- 0.1& \SI{}{day^{-1}}  & \cite{he2020temporal,pan2020time,Ling2020Persistence} \\
$\gamma_\ICU$       & Recovery rate from \ICU               & 0.13      & 0.08 -- 0.2& \SI{}{day^{-1}}  &  \cite{Levin2020,Linden2020DAE,salje2020estimating,bauer2021relaxing} \\
$\delta$            & Av. hospitalization rate $I\to\ICU$       & 0.0019    & $10^{-5}$ -- 0.007 & \SI{}{day^{-1}}  &  \cite{Levin2020,Linden2020DAE,salje2020estimating,bauer2021relaxing} \\
$\theta$       & Av. death rate& $5.4\,10^{-4}$    & $2\,10^{-6}$ -- 0.005 & \SI{}{day^{-1}}  &  \cite{Levin2020,Linden2020DAE,salje2020estimating,bauer2021relaxing} \\
$\theta_\ICU$  & Av. \ICU death rate   & 0.0975   & 0.088 -- 0.100 & \SI{}{day^{-1}}  &  \cite{Levin2020,Linden2020DAE,salje2020estimating,bauer2021relaxing} \\
$\alpha$            & Sensitivity of the population to ICU occupancy   & --     & -- & \SI{}{day^{-1}}  &  Estimated \\
$\Omega_v^{\rm base}$  & Waning imm. rate (base, vaccination) & $\frac{2/3}{360}$  & -- & \SI{}{day^{-1}}   & \cite{goldberg2021waning,saure2021dynamic} \\
$\Omega_n^{\rm base}$  & Waning imm. rate (base, natural)     & $\frac{1}{360}$   & -- & \SI{}{day^{-1}}   &  \cite{turner2021sars,wang2021naturally} \\ 
$\mu$  & Sensitivity to seasonality                         & 0.267   & 0.141--0.365 & --  & \cite{Gavenciak2021seasonality} \\
$d_\mu$  & Day with the strongest effect on seasonality     & 0   & -- & \SI{}{day}  & \cite{Gavenciak2021seasonality} \\\midrule
$d_0$    & Day when the time series starts                   & 240   & -- & \SI{}{day}  & \cite{Gavenciak2021seasonality} \\\midrule
$\phi_0, \varphi_0$ & Administration rate (first-time and booster doses resp.)     & 0.0025    &   --       & \SI{}{day^{-1}}        &  \cite{bauer2021relaxing}\\
$\chi_0, \chi_1$ & Fraction of the population refusing vaccine (first and booster resp.). & 0.1, 0.2    &   --       & --        &  \cite{betsch2020germany}\\
$u_{\rm base}$ & Base acceptance of first dose & 0.5    &   --       & --        &  \cite{wouters2021challenges}\\
$\Psi$ & Influx of infections & 1    &   --       & People/day        & assumed\\\bottomrule
\end{tabular}%
\end{table}

\begin{table}[htp]\caption{Model variables.}
\label{tab:Variables}
\hspace*{-1cm}
\centering
\begin{tabular}{l p{4cm} l  p{9cm} }\toprule
Variable & Meaning & Units & Explanation\\\midrule
$M$               & Population size       & \SI{}{People} & Default value: 1000000\\
$S$ & Susceptible pool & \SI{}{People} & Non-infected people that may acquire the virus.  \\
$V$ & Vaccinated pool & \SI{}{People} & Non-infected, vaccinated people. Less likely to be infected or develop severe symptoms  \\
$W$ & Waned immunity pool & \SI{}{People} & Non-infected people whose immunity (vaccine-induced or natural) has already waned, thus may acquire the virus.  \\
$E$ & Exposed pool   & \SI{}{People} & People exposed to the virus. \\
$\EB$ & Exposed pool (breakthrough infection)    & \SI{}{People} & People exposed to the virus (breakthrough infection).\\
$I$ & Infectious pool   & \SI{}{People} & Infectious people. \\
$\IB$ & Infectious pool (breakthrough infection)    & \SI{}{People} & Infectious people (breakthrough infection).\\
$\ICU$ & Hospitalized (total)    & \SI{}{People} & Hospitalized people.\\
$R$ & Recovered (total)    & \SI{}{People} & Recovered people (naturally or after requiring intensive care).\\
$H$ & Av. ICU occupancy   & \SI{}{People} & Auxiliary variable measuring the average ICU occupancy.\\
$u^{\rm current}, w^{\rm current}$ & Vaccinated individuals, independent of the compartment  & \SI{}{People} & Integral over the vaccination rates\\
$\RtH$              & Reproduction number (gross)      & \SI{}{-}        & Eq.~\ref{eq:Rt} \\
$N$ & New infections (Total) & \SI{}{cases\, day^{-1}} & $N = \gamma R_t(\ICU)\frac{\Ieff}{M}\left(S+W + \left(1-\fracImmun\right)V\right)$.\\
\avProtection           & Effective vaccine efficacy against severe course  & \SI{}{cases\, day^{-1}} & $\avProtection = 1-\frac{1-\kappa_{\rm obs}}{1-\fracImmun}$\\ 
$\Gamma$ & Seasonal var. of SARS-CoV-2 infectiousness&  \SI{}{-} & Eq.~\ref{eq:seasonality}. \\
$c_v, c_n$  & Inverse incidence at which waning immunity is halved & \SI{}{People}   & Eq.~\ref{eq: waningparameter} \\
$\phi(t), \varphi(t)$ & Administration rate of first-time and refreshing vaccine doses (resp.)  &  \SI{}{doses/day}        &  Eq.~\ref{eq: phi}, \ref{eq:varphi}\\
\bottomrule
\end{tabular}%
\end{table}

\subsubsection{Initial conditions}
\label{sec:initials}

Initial conditions for Fig. \ref{fig:Figure_1} were vaguely inspired by the situation in Germany as of September 1st, 2021. The population size in our model is set to $M=10^6$ individuals. Let $x$ be the vector collecting the variables of all different compartments:

\begin{equation}
    x = [S, V, W, E, E^B, I, I^B, \ICU, R, D, u^{\rm current}, w^{\rm current}, \avICU]
\end{equation}

In that way, $\sum_{i\leq 10}x_i = M$ because $u^{\rm current}$ and $, w^{\rm current}$ are counted independent of their compartment and $\avICU$ is a measure for past ICU. The NPI-related reproduction number $\RtBase$ is initially set corresponding to one of our three different scenarios and then increased linearly to 5 in March 2022, which corresponds to the natural reproduction number of the virus reduced only slightly by e.g., improved hygiene that is kept in place. 

The precise values used for Fig. \ref{fig:Figure_1} can be found in Tab. \ref{tab:initial_conditions_Fig1}.


Note that the $V$ and $R$ compartment overlap which is why their sum is NOT equivalent to the immunized fraction of the population. Firstly, a waned fraction is subtracted from $V$ and $R$. Secondly, we estimate the overlap by assuming that the majority of the population was infected before being vaccinated in case they fall in both categories. That way, recovered people have the same probability to also have a vaccine and we can calculate the overlap as the product $RV$. We subtract the overlay from $V$ instead of $R$ because of longer lasting natural immunity.  


We calculate the initial conditions for the exposed and infected compartments by first estimating $E+\EB$ as $\frac{1}{\rho}$ times the daily new cases, taken from official data as of September 1st, 2021. This should not in any way try to predict the precise case numbers in the future but only serve as initial conditions that are suited for the system. The infectious groups $I+\IB$ are estimated as $\frac{1}{\gamma+\delta+\theta}$ times the daily new cases. To find the fraction of breakthrough infections among all infected individuals we calculate their fraction as $\frac{V(1-\eta)+W}{S+W+V(1-\eta)}$ and build up the compartments $E$, $\EB$, $I$ and $\IB$ accordingly. We estimate the initial condition for $W$ as the sum of $W_V+W_R$ where $W_V$ is the fraction of people whose vaccine-induced immunity has waned and $W_R$ is the fraction of people whose natural immunity has waned. Both are estimated as 10\% of the vaccinated and recovered respectively.

\begin{table}[htp]\caption{\textbf{Initial conditions for Fig. \ref{fig:Figure_1}} Each compartment is normalized to the population size $M$ and represented here in either percentages or total numbers. }
\label{tab:initial_conditions_Fig1}
\centering
\begin{tabular}{lllllllllllll}\toprule
 $S$  [\%] & $V$  [\%] & $W$  [\%] & $E$ &  $E^B$ & $I$ & $I^B$ & $\ICU$  & $R$  [\%] & D & $u^{\rm current}$  [\%] &  $w^{\rm current}$  [\%]  \\\midrule
30.6  & 43.2 & 8 & 333 & 66 & 813 & 163& 10& 18 & 0& 60 & 0   \\
\bottomrule
\end{tabular}%
\end{table}

For initial conditions that represent different situations, considering that other countries have different levels of vaccinated, see sec. \ref{sec: othercountries}

\subsection{Vaccination effects} \label{sec:vaccinationeffects}

Our model includes the effect of vaccination, where vaccines are for simplicity administered with a single-dosage delivery scheme. There is some evidence that the vaccines partially prevent the infection with and transmission of the disease \cite{mallapaty_can_2021, hall_effectiveness_2021}. Our model incorporates both the effectiveness against infection and against a severe course of the disease following a 'leaky' scheme, i.e., vaccinated individuals have smaller chances to be infected by a factor of $\left(1-\eta\right)$, and those with a breakthrough infection or waned immunity have a lower probability of going to ICU by a factor of $\left(1-\kappa\right)$ than unvaccinated individuals, where $\kappa$ can be obtained from 
\begin{equation}\label{eq:kappa}
    (1-\eta)(1-\kappa)=(1-\kappa_\mathrm{obs}),
\end{equation}
with $\kappa_\mathrm{obs}$ denoting the full protection from severe disease as observed in studies.
Furthermore, we assume breakthrough infections carry a lower viral load and are thus less infectious by a factor of $\sigma$ \cite{harris2021impact}. The infection rate depends on the sum of the infected compartment, the breakthrough infected compartment with a lower viral load and an external influx. It can be expressed via the effective incidence $I_{\rm eff} =\left(I+\sigma\IB+\Psi\right)$. All parameters and values are listed in~\tabref{tab:Parametros}. Note that these parameters are to be understood as averages across vaccine types.

\subsection{Vaccine uptake}

\begin{figure}[!h]
    \centering
    \includegraphics[width=4.5in]{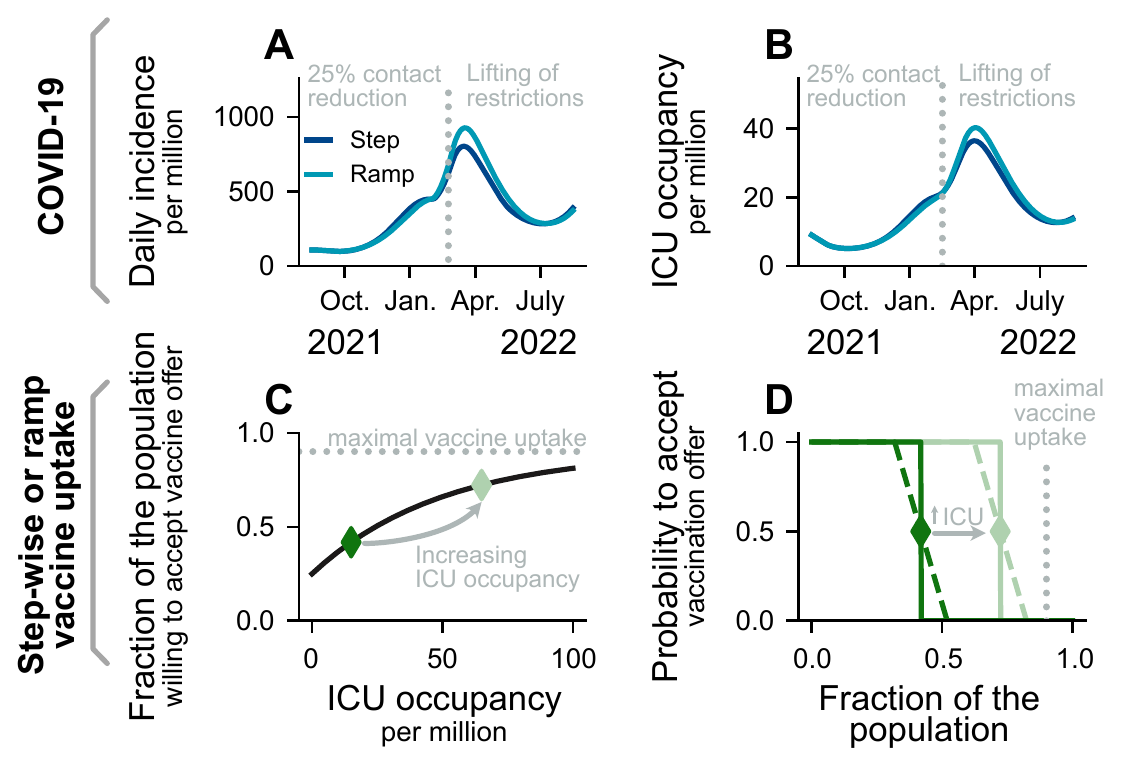}
    \caption{%
    \textbf{Modeling choices for the opinion-epidemic feedback loop, modulating vaccine uptake and contacts, affects case development in comparable pandemic situations.}
    \textbf{A, B:} Assuming that individuals react protectively to the information they receive from the pandemic, self-regulation of contacts and vaccine uptake could stabilize disease spread. The model approach to this affects disease spread. \textbf{C, D:} Assuming that vaccine uptake and willingness are decoupled, we can represent vaccine uptake from an on-off perspective happening when vaccine willingness (partially modulated by ICU occupancy) is higher than the current uptake. In this setting, vaccination can be a step function or a ramp centered where vaccine uptake meets the vaccine willingness.
            }
    \label{fig:Figure_S2}
\end{figure}

Vaccine dynamics is an important aspect of our model because it has a strong influence on infection dynamics due to reduced transmissibility. Incorporating willingness to be vaccinated into our model requires making a decision on how vaccines are administered. Vaccine uptake is described by two different functions, one for susceptible individuals ($\phi$) and one for individuals whose immunity has waned ($\varphi$). The idea is to vaccinate only if willingness for vaccine uptake is larger than the fraction of already vaccinated. We compare a step-wise approach for this transition described in \ref{sec:stepwiseapproach} with a ramping approach described in \ref{sec:rampingapproach} that is used for Fig. \ref{fig:Figure_1}. A comparison between the outcomes of the two methods is shown in Fig. \ref{fig:Figure_S2}. 

\subsubsection{Step-wise approach}
\label{sec:stepwiseapproach}
In this approach, we use functions that represent the willingness to be vaccinated in dependence on the ICU occupancy. If the group of individuals who are willing to be vaccinated with a first dose ($u^{\rm willing}$) is larger than the group of already vaccinated ($u^{\rm current}$), vaccinations are carried out at a rate proportional to the difference of the two, or at a maximum administration rate $\phi_0$, depending on which one is lower:

\begin{align}
    \phi(\avICU_u) & = \left\{ \begin{array}{ll} \dis \min\left\{\phi_0,u^{\rm willing}(\avICU_u)-\frac{1}{M}u^{\rm current}\right\}  & \text{if }  u^{\rm willing}(\avICU_u)\geq \frac{1}{M}u^{\rm current}, \\                       \dis 0 &                   \text{else.}\end{array} \right. 
    \label{eq: phi}
\end{align}

$\avICU$ is a function dependent on the past development of the ICU occupancy as discussed in \ref{sec:memoryICU}. The fraction of people who are willing to be vaccinated for the first time can shift between a minimum and a maximum value ($u_{\rm base}$ and $u_{\rm max}= 1-\chi_0$), representing the general observed acceptance for the first dose and people who are strictly opposed to vaccines or cannot be immunized because of age or other preconditions respectively. Willingness to be vaccinated depends on perceived danger, which the ICU occupancy is a suitable measure for. The willingness is then represented by 

\begin{align}
     u^{\rm willing} & = u_{\rm base} + \left(u_{\rm max} -u_{\rm base}\right)\left(1-\exp\left(-\alpha_u \avICU_u \right)\right)\,.
\end{align}

Willingness to accept booster doses is modeled in a similar way, without a base willingness:

\begin{align}
    w^{\rm willing} & =\left(1-\chi_1\right)\left(1-\exp\left(-\alpha_w \avICU_w \right)\right)\,.
\end{align}

If the number of people willing to be vaccinated with a booster dose is larger than the number of people that already received one, vaccinations are carried out from the waned compartment $W$ to compartment $V$ at a rate

\begin{align}
     \varphi(\avICU_w) & = \left\{ \begin{array}{ll} \dis \min\left\{\varphi_0,w^{\rm willing}(\avICU_w)-\frac{1}{M}w^{\rm current}\right\}  & \text{if }  w^{\rm willing}(\avICU_w)\geq \frac{1}{M}w^{\rm current}, \\                       \dis 0 &                   \text{else.}\end{array} \right.
     \label{eq:varphi}
\end{align}

\subsubsection{Ramping approach}
\label{sec:rampingapproach}
In our first step-wise approach vaccinations are only carried out if more people are willing to be vaccinated than there are currently vaccinated. This hard transition might not be realistic because it can be assumed that in a real-world scenario the transition would be smooth, leading to vaccinations being carried out over a longer time frame and not abrupt. We can twist our first approach to incorporate this effect by replacing the step function $\phi(\avICU_u)$ by

\begin{align}
    \phi(\avICU_u) & = \left\{ \begin{array}{ll} \dis 0 &    \text{if }  u^{\rm willing}(\avICU_u) \leq \frac{1}{M}u^{\rm current}-\epsilon \\                       
    \dis \phi_0 &      \text{if }  u^{\rm willing}(\avICU_u) \geq \frac{1}{M}u^{\rm current}+\epsilon   \\
    \dis \frac{\phi_0}{2\epsilon}\left(u^{\rm willing}(\avICU_u)-\frac{1}{M}u^{\rm current}+\epsilon\right) & \text{else. }  \end{array} \right. 
    \label{eq: rampphi}
\end{align}
For booster doses, the replacement $\phi \leftrightarrow\varphi$ and $u\leftrightarrow w$ has to be made.
This corresponds to a linear increase with a slope dependent on $\epsilon$, where vaccinations are started to be carried out when vaccination willingness is larger than the current vaccinated minus $\epsilon$. That way, $\epsilon$ is a measure for the smoothness of the transition between the state where vaccinations are carried out and the state where they are not. 

\subsubsection{Tracking of vaccinated individuals}

In our first two approaches, vaccination rates between the susceptible ($S$) and waned ($W$) compartment depend on the difference between willingness for vaccination uptake and the currently vaccinated. Thus, it is necessary to keep track of how many people received a first and booster dose respectively. Implementing this is achieved by integrating over the vaccination rates. It translates into two additional differential equations:

\begin{equation}
    \frac{d}{dt}u^{\rm current} = M\phi(\avICU) \hspace{0.5cm} \text{and} \hspace{0.5cm}  \frac{d}{dt}w^{\rm current} = M\varphi(\avICU)\,.
\end{equation}
The initial conditions for $u^{\rm current}$ and $w^{\rm current}$ are chosen according to the initial conditions of $V$ and $W$.

\subsection{Assessment of the sensitivity to ICU occupancy}
\label{sec:assessmentAlphas}
The opinion dynamics in our model depend on the parameters $\alpha_R$, $\alpha_u$, and $\alpha_w$. To get an estimate for their magnitude we look at estimated incidences that cause a change in the stability of the system due to changes in behavior. The effective reproduction number $R_t^{\rm eff}$ is defined as the reproduction number times the fraction of the population that is susceptible: 
\begin{equation}
   R_t^{\rm eff}  \approx  \frac{S+W}{M}R_t(\avICU_R) + \frac{V}{M}(1-\fracImmun) R_t(\avICU_R) .
\end{equation}
If $R_t^{\rm eff} = 1$ the system is at an equilibrium which means that the incidence is constant. Imposing equilibrium conditions at time $t=t_{\rm eq}$, we can obtain an equation ruling the balance between all stabilizing and destabilizing contributions:

\begin{equation}
    \exp\left(-\alpha_R\avICU_{R,\rm eq} \right)\RtBase \Gamma(t_{\rm eq})\left(\frac{S_{\rm eq}+W_{\rm eq}}{M} + \frac{V_{\rm eq}}{M} (1-\fracImmun)  \right) \overset{!}{=} 1
\end{equation}

Thus, we can estimate $\alpha_R$ as:
\begin{equation}
    \alpha_R \approx \frac{1}{\avICU_{R,\rm eq}}\ln\left(\RtBase \Gamma(t_{\rm eq}) \frac{{S_{\rm eq}+W_{\rm eq}+(1-\fracImmun)V_{\rm eq}}}{M}\right).
\end{equation}

Solving for $\alpha_R$ we need to assume at which ICU occupancy the behavioral changes are strong enough to lead to a tipping of the system. This method is usable to obtain the right orders of magnitude. Setting $\RtBase=4$ and $\Gamma(t_{eq})=1$ corresponds to an equilibrium situation in winter with mild restrictions. Estimating the vulnerable fraction of the population at that point to be a half, we can estimate $\alpha_R\approx\frac{\text{ln}2}{H_{R,\rm eq}}\approx 0.1$. For $\alpha_u$ and $\alpha_w$ we assume that the decision to get vaccinated does not require as much self discipline as contact reduction and estimate them as twice and three times larger than $\alpha_R$ respectively: $\alpha_u\approx 0.2$, $\alpha_w\approx 0.3$.

\section{Different initial conditions}
\label{sec: othercountries}

Our results were obtained using initial conditions that roughly reflect the situation in Germany around 1. September 2021. In the following we simulate scenarios with different initial conditions (inspired by other countries). Of particular interest are initial conditions with different vaccination fractions among the population. We expect that these differences shape the extent of the winter dilemma. We distinguish between four sets of example initial conditions: 

\begin{itemize}
    \item \textbf{Example 1:} Low vaccination.
    \item \textbf{Example 2:} Moderate vaccination.
    \item \textbf{Example 3:} High vaccination. 
    \item \textbf{Example 4:} Very High vaccination.
\end{itemize}

\begin{figure}[!ht]
    \centering
    \includegraphics[width=6in]{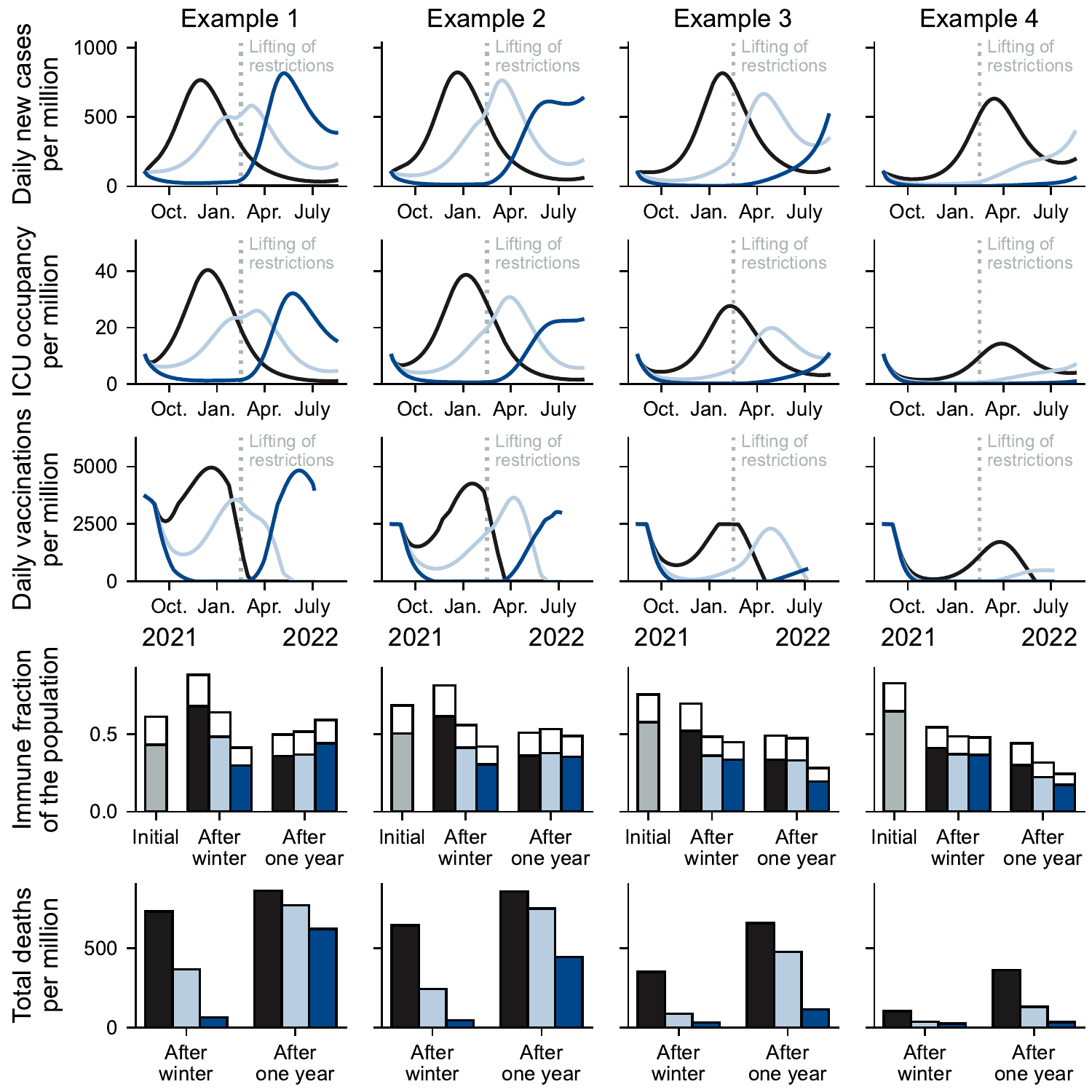}
    \caption{%
        \textbf{Differences between the levels of immunity and their nature will determine the extent of the winter dilemma across countries.} Considering the same scenarios of Fig.~\ref{fig:Figure_1} in the main text (0, 25\% or 50\% of restrictions in winter), here we study the effect of different initial conditions in immunity on the extent of the winter dilemma. Across scenarios, we see that countries with higher initial immunity will have less severe waves. However, considering waning immunity and low vaccination rates, their preparedness for the 2022 winter wave would be lower. While these results can hold in the short-term, we expect them to be affected by differences in i) the definition of intensive care and the hospital resources, ii) population's risk perception and sensitivity to ICU occupancy, compliance to NPIs, age-stratified vaccine uptake, and degree of solidarity, and iii) contact structure and intensity.Thus, the above highlights the need for research in this direction. Note that in the fourth row, the upper white fraction (immune by infection) also includes those who had been infected and then also vaccinated. Furthermore, it is reduced by the waned fraction of the population, for further explanation see Sec. \ref{sec:initials}). 
        }
    \label{fig:Figure_S3}
\end{figure}

All other initial conditions are calculated as described in \ref{sec:initials}, with daily new cases per million set to 100 and the ICU occupancy set to 10. Effectively, this translates to different initial conditions in all compartments except $\ICU$ (exposed and infectious groups become different due to different vaccination levels). Fig.~\ref{fig:Figure_S3} shows the comparison between the four examples.       

\begin{table}[htp]\caption{\textbf{Country-dependent parameters for Fig.~\ref{fig:Figure_S3}.} Vaccinated fraction  $\dis\frac{V}{M}$ represents all people who received a vaccine.
$\dis\frac{R}{M}$ represents all people with naturally acquired immunity, including those with additional vaccination, \textit{which is why both fractions can add up to more than 100\%.} The overlap between $\dis \frac{V}{M}$ and $\dis \frac{R}{M}$ (see Sec. \ref{sec:initials}) is estimated and subtracted from $\dis \frac{V}{M}$ because of longer lasting natural immunity. 
Waned fractions $\dis \frac{W_V}{M}$ (waned vaccine-induced immunity) and  $\dis \frac{W_R}{M}$ (waned natural immunity) are assumed to be 10\% of vaccinated and recovered respectively. }
\label{tab:countryparameters}
\centering
\begin{tabular}{l  cccc}\toprule
Example &  $\dis\frac{V}{M} [\%]$ \cite{owidcoronavirus} & $\dis\frac{R}{M} [\%]$ \cite{arora2021serotracker} & $\dis\frac{W_V}{M} [\%]$ & $\dis\frac{W_R}{M} [\%]$   \\\midrule
1: "low vaccination "             & $60$ & $20$ & 6 & 2  \\
2: "moderate vaccination "        & $70$ & $20$ & 7 & 2\\
3: "high vaccination"  & $80$ & $20$ & 8 & 2 \\
4: "very high vaccination"           & $90$ & $20$ & 9 & 2  \\
\bottomrule     
\end{tabular}%
\end{table}

\end{document}